\documentclass[11pt,english,a4paper]{article}
\usepackage{natbib}
\usepackage{rotating,placeins}
\usepackage{babel}
\usepackage{geometry}
\usepackage{fancyhdr}
\usepackage{lscape,epsfig,amssymb,amsmath}
\usepackage[]{subfigure}

\newcommand{\Sec}[1]{Sec~\ref{sec:#1}}
\newcommand{\Fig}[1]{Fig~\ref{fig:#1}}
\newcommand{\Eq}[1]{Eq~(\ref{eq:#1})}

\newcommand{\Hr}{$H_{100}\,$}

\newcommand{\Ur}{$U_{100}\,$}

\title{Wind and Wave Extremes over the World Oceans From Very 
       Large Forecast Ensembles}

\author{{\O}yvind Breivik\footnote{Final version published as Breivik, {\O}, O~J
Aarnes, S Abdalla, J-R Bidlot (2013). \emph{Wind and Wave Extremes over the World Oceans From Very Large
 Forecast Ensembles}, in Proceedings of the 13th International Workshop on Wave 
 Hindcasting, Banff, Canada }
\thanks{Corresponding author. E-mail: \texttt{oyvind.breivik@ecmwf.int}}
\thanks{European Centre for Medium-Range Weather Forecasts (ECMWF), Shinfield Park, Reading, RG2 9AX, 
 United Kingdom}, 
\and Ole Johan Aarnes\thanks{Norwegian Meteorological Institute}
\and Saleh Abdalla\footnotemark[3]
\and Jean-Raymond Bidlot\footnotemark[3]}

\begin{document}

\maketitle                                                                                                          

\abstract{
Global return value estimates of significant wave height and 10-m neutral wind
speed are estimated from very large aggregations of archived ECMWF ensemble
forecasts at +240-h lead time from the period 2003-2012.  The upper percentiles
are found to match ENVISAT wind speed better than ERA-Interim (ERA-I), which
tends to be biased low. The return estimates are significantly higher for both
wind speed and wave height in the extratropics and the subtropics than what
is found from ERA-I, but lower than what is reported by \citet{caires05b}
and \citet{vinoth11}.  The highest discrepancies between ERA-I and ENS240
are found in the hurricane-prone areas, suggesting that the ensemble comes
closer than ERA-I in capturing the intensity of tropical cyclones. The width
of the confidence intervals are typically reduced by 70\% due to the size
of the data sets. Finally, non-parametric estimates of return values were
computed from the tail of the distribution.  These direct return estimates
compare very well with Generalized Pareto estimates.}

\newpage

\section{Introduction}
Return value estimates of significant wave height and 10-m wind speed
over the oceans are fundamental to risk assessments.  Long observational
records are scarce, making global return value estimates impossible,
with the exception of altimeters \citealt{vinoth11}, which to date
represent rather short and heterogeneous time series.  Reanalyses and
hindcasts \citep{kalnay96,upp05,weisse07,dee11,rei11,wang12} serve as
proxies for observations where there are none. But even long reanalyses
will normally be substantially shorter than the return period sought,
leaving wide error bars on the return values computed from them
\citep{wang01,wang02,caires05b,sterl05,bre09}.

The Integrated Forecast System (IFS) of the European Centre for Medium-Range
Weather Forecasts (ECMWF) has produced daily ensemble forecasts (ENS) since 1992
\citep{mol96,buizza07,hagedorn08} and has been coupled to the WAM wave model
since 1998 \citep{jan04,janssen08}.  Even though the ECMWF forecast skill has
steadily been improving over the years \citep{richardson10}, +240-h lead time 
ensemble
forecasts of wind and waves still tend to be weakly correlated, and their upper
percentiles are virtually uncorrelated \citep{bre13b}. Such weak correlations
are in fact a necessity when utilizing ensemble forecasts for extreme value
estimation since the data must be independent and identically distributed.  This
may appear as something of a paradox since it means the forecasts can only be
used for estimating return values if the skill is low.

Aggregating large amounts of virtually uncorrelated ensemble forecasts
to estimate 100-year return values of wave height was first explored
by \citet{bre13b}. They found that the estimates matched observed upper
percentiles well in the Norwegian Sea. The results were also found to agree
fairly well with estimates based on the high-resolution hindcast NORA10
\citep{rei11,aar12,semedo13}. In this study we extend the analysis to 10-m wind speed
and compare global 100-yr return values with return values computed from the
ERA-I reanalysis. We also evaluate the upper percentiles against buoy
measurements and ENVISAT altimeter wind speed observations.

\section{Methods}
\label{sec:methods}
ENS forecasts at +240 h lead time were interpolated onto a regular $1\times
1^\circ$ grid. All forecasts (two per day, 00 and 12 UTC) between March 2003
and March 2012 were used. Model values were collocated with buoy observation
locations using a bilinear interpolation.  ERA-I fields (1979-2012) were
interpolated onto the same grid as ENS.  The current spectral truncation of
ENS is T639 for the atmospheric model, corresponding to approximately 32 km,
with the wave model run at approximately 55 km. Previous model cycles had
coarser resolution (see Fig 1 by \citealt{bre13b}).

The two daily ensembles of +240-h forecasts from 50 perturbed ensemble members
plus the unperturbed control member aggregated over 9 years amount to
\begin{equation}
    51\, \mathrm{members} \times\,2\, \mathrm{daily\,forecasts} \times 6
    \mathrm{h}\,
    \times\,9\
    = 229.5 \,\mathrm{yr}
\end{equation}
under the assumption that each forecast is representative of a six-hour 
interval \citep{bre13b}.

The 10-m neutral wind speed was extracted for both ENS and ERA-I. This is the
field used to force the wave model and is thus consistent with the significant
wave height fields investigated.  Grid points with less than 80\% ice-free
forecasts were excluded from the analysis when computing the return values for
significant wave height.  No such filtering was required for the wind speed.

\begin{figure}[h]
\begin{center}
\includegraphics[scale=0.35]{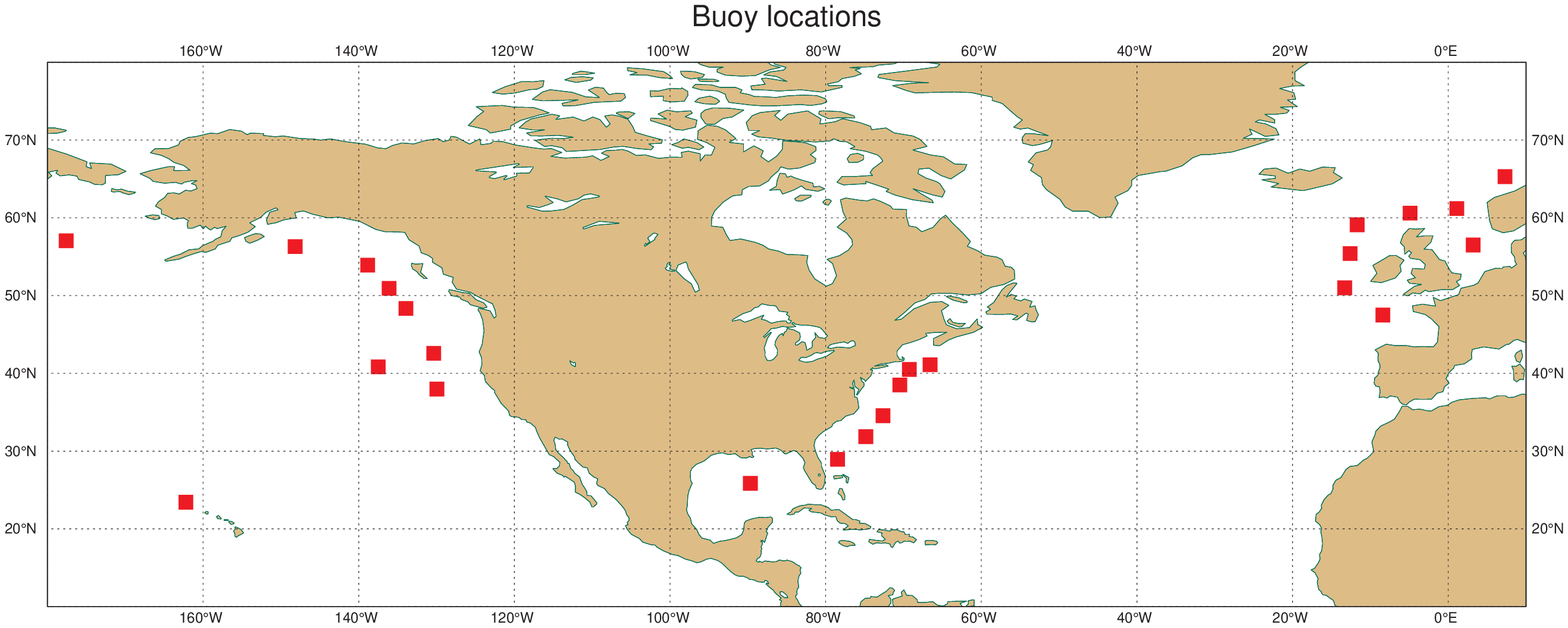}
\caption{
24 wind and wave-measuring buoys were used to assess the upper-percentile
climatology of the ENS forecasts.}
\label{fig:buoys} 
\end{center} 
\end{figure}

The wave height from 24 buoys (see \Fig{buoys}), averaged over a period
of $\pm 2$ hours, were collocated with model data for verification
\citep{bidlot02,bre13b}.  Only wind and wave measuring buoys in deep water
($>70$ m) were selected since coarse resolution forecasts are ill-suited
for simulating near-shore conditions.  Furthermore, ENVISAT RA2 altimeter
observations of wind speed and wave height were averaged into along-track
``super-observations'' of comparable resolution to the WAM grid
\citep{abdalla04}. This procedure
makes data and model values more comparable, which is important when assessing
the upper percentiles of the model climatology for wind and waves.

The return estimates from ENS were found using the Generalized Pareto (GP)
distribution for data exceeding a threshold $u$ such that $y = X_i - u$,
$y > 0$. The GP distribution reads as \citep{col01}
\begin{equation}
   H(y) = 1 - \left(1+\frac{\xi y}{\tilde{\sigma}}^{-1/\xi}\right).
   \label{eq:gp}
\end{equation}
Here $\xi$ is the shape parameter and $\tilde{\sigma}$ is a scale parameter.
In the limit as $\xi \to 0$ \Eq{gp} tends to the exponential distribution.
The threshold was set to the 1000 highest forecasts, corresponding to the
$99.7\%$ percentile. Since ensemble forecasts are assumed uncorrelated, all
data points exceeding the threshold were used. For ERA-I, the threshold was
also set to 99.7\% and a standard peaks-over-threshold (POT) technique where
data must be separated by 48 h was used to assure that points are independent
and identically distributed \citep{mathiesen94,lop00,col01,aar12,bre13b}.
Confidence intervals were estimated using the \emph{Delta method}
(\citealt{col01}, p 33).

Extreme value distributions are parametric estimates of the underlying
distribution of the theoretical maxima based on modelled or observed
values, usually in the form of one or more continuous time series with a
fixed temporal resolution. However, under the assumptions that a forecast
represents a temporal period we may convert
our collections of nearly independent ensemble forecasts into an equivalent
time series which is significantly longer than 100 years (229 years). In
other words, we can make non-parametric direct return estimates (DRE) from
the ensemble of the 100-yr return value, $H_{100}^\mathrm{DRE}$, without
invoking an extreme value distribution.  However, some care has to be taken
when interpreting the upper percentiles of even quite large data sets since
the nature of extremes is such that the true return value for a given extreme
value distribution will only have a certain probability of appearing in
any given period. The probability of exceeding the 100-year return value
in any given 100-year period is about 63\% for the Gumbel distribution.
Complementary, there is still a certain probability (${\sim}10$\%) that the
100-year return value does not appear in our data set.
The DRE method interpolates the tail of the cumulative distribution,
For a data set of ${\sim}229$ years 
a linear interpolation between $X_{(2)}$ and $X_{(3)}$
(the second and third highest values in an ordered series) yields the
following weighting,
\begin{equation}
   r_{100}^\mathrm{DRE} = 0.67X_{(2)}+0.33X_{(3)},
\end{equation}
where $r_{100}$ is the 100-yr return value.

\section{Results and Discussion}
\begin{figure}[h]
\begin{center}
(a)\includegraphics[scale=0.5]{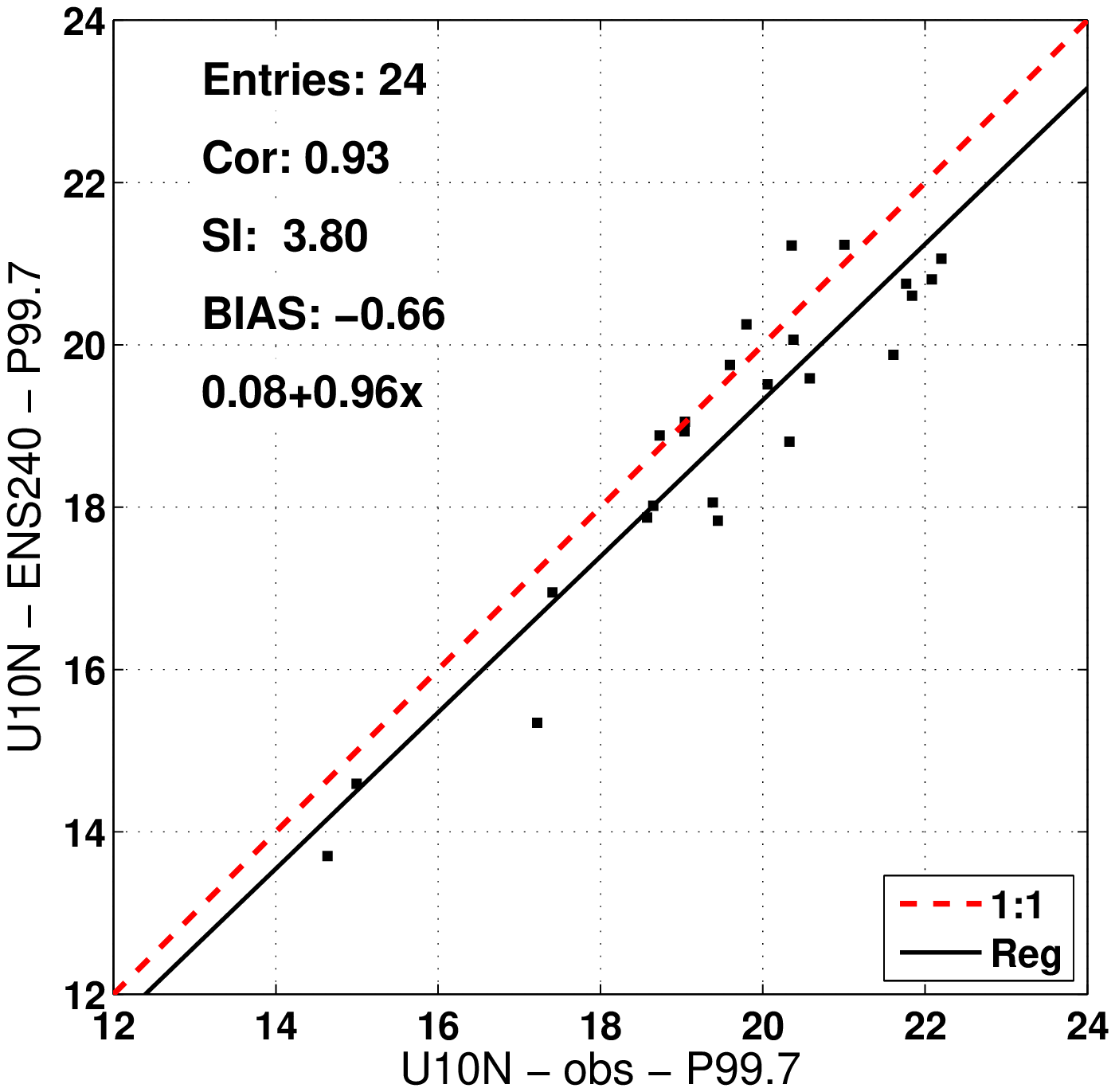}\\
(b)\includegraphics[scale=0.5]{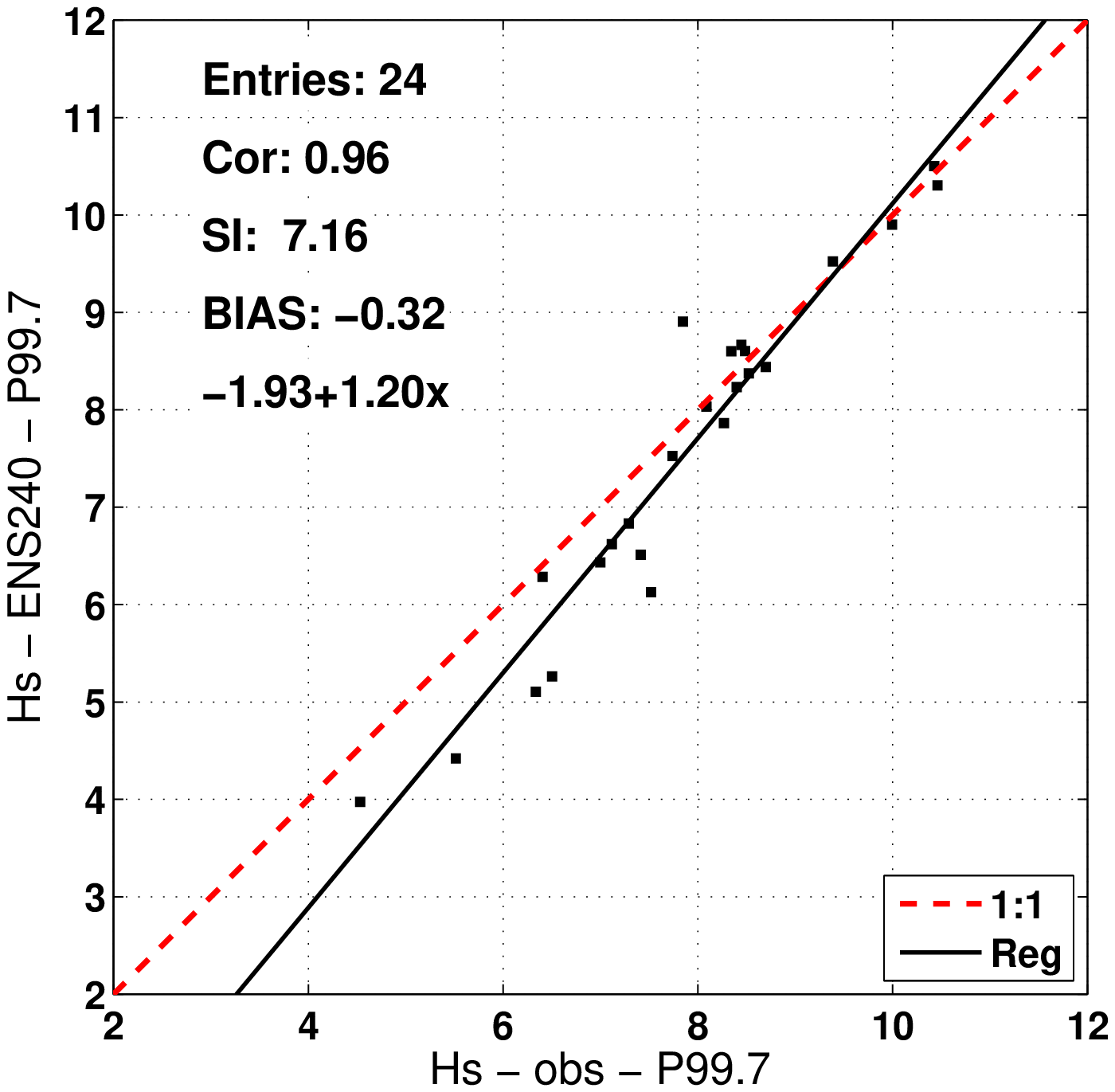}\\
\caption{
Panel a: Observed \emph{v} modelled 99.7 percentiles ($P_{99.7}$) of 10-m neutral wind 
speed [m s$^{-1}$]. Panel b: Significant wave height [m]. 
}
\label{fig:P99} 
\end{center} 
\end{figure}

\begin{figure}[h]
\begin{center}
(a)\includegraphics[scale=0.5]{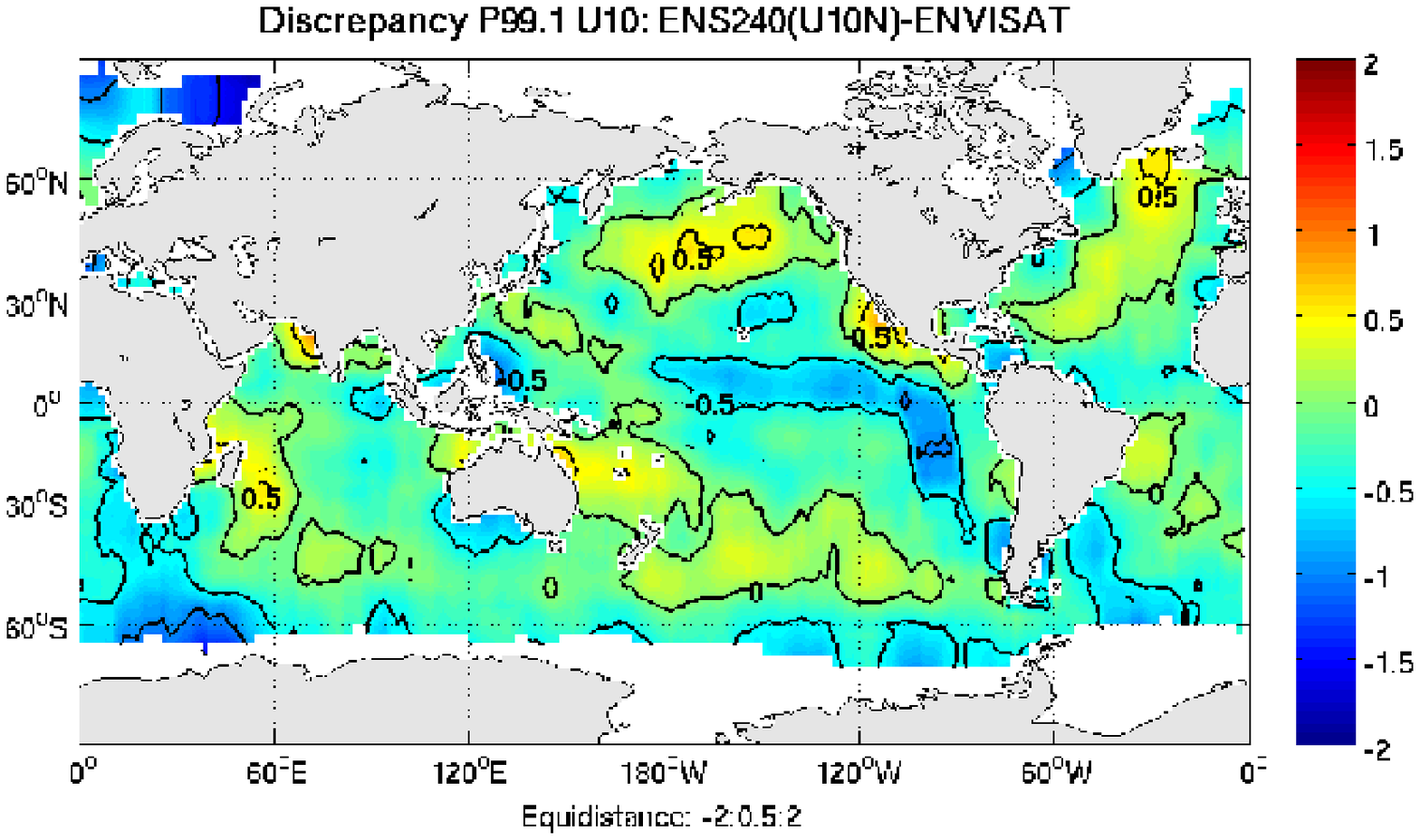}\\
(b)\includegraphics[scale=0.5]{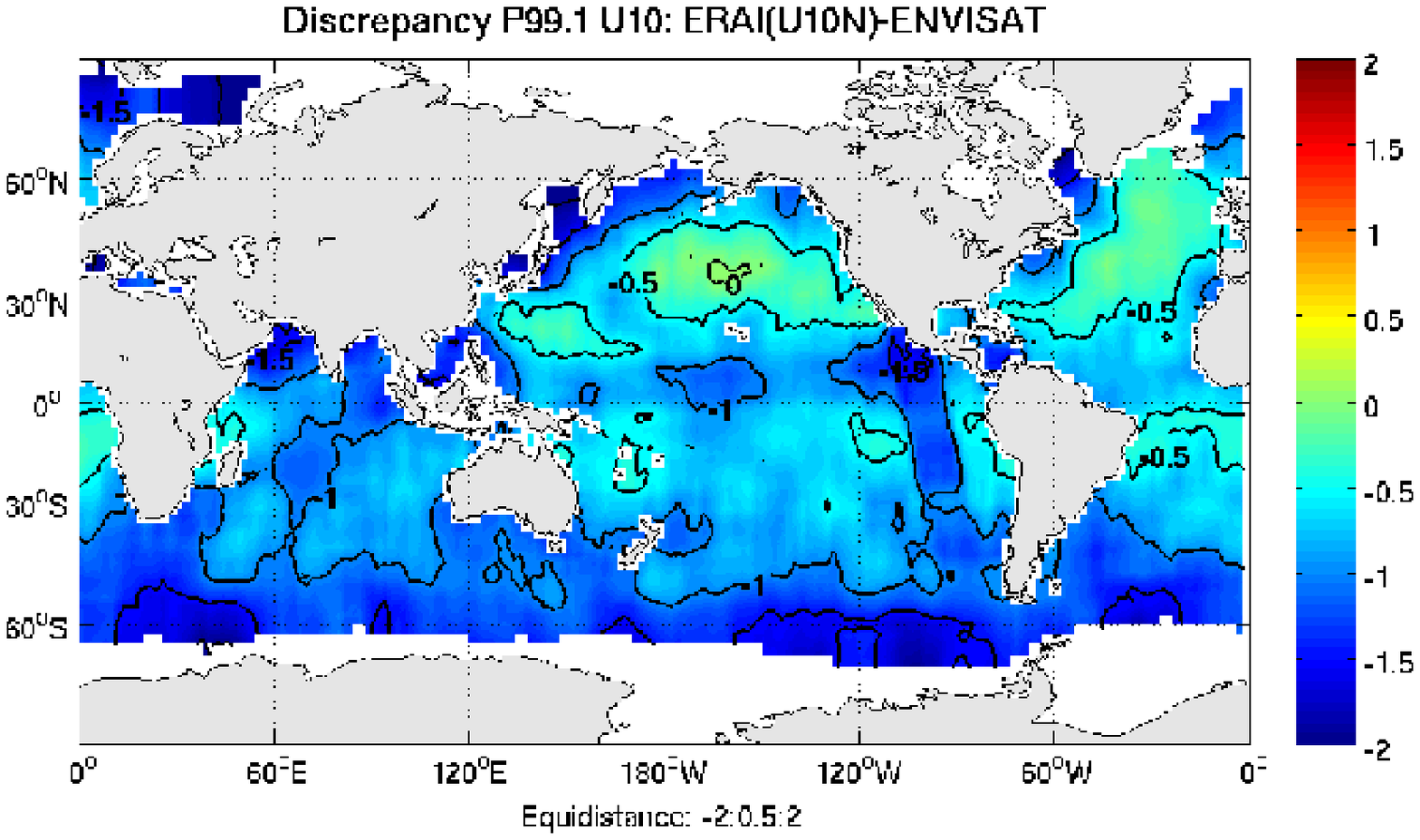}
\caption{
Panel a: The difference between the ENVISAT altimeter 99.1 percentile ($P_{99.1}$)
10-m wind speed (2002-2012) and ENS240 neutral 10-m wind speed  [m s$^{-1}$]
(positive when
ENS240 is larger than ENVISAT). Panel b: Difference between
ERA-I and ENVISAT wind speed.
The differences between ERA-I and ENVISAT at $P_{99.1}$ are generally larger
(ERA-I biased low) than for ENS240.
}
\label{fig:envisat} 
\end{center} 
\end{figure}

The tail of the forecast distribution should closely resemble the observed
distribution. \Fig{P99} shows good agreement at the 99.7 percentile level for
significant wave height and wind speed throughout the northern hemisphere
oceans (buoy locations shown in \Fig{buoys}).  \Fig{envisat} compares the
$P_{99.1}$ of ENVISAT altimeter wind speed. As can be seen ENS240 has less
bias than ERA-I at the tail of the distribution.

\begin{figure}[h]
\begin{center}
(a)\includegraphics[scale=0.5]{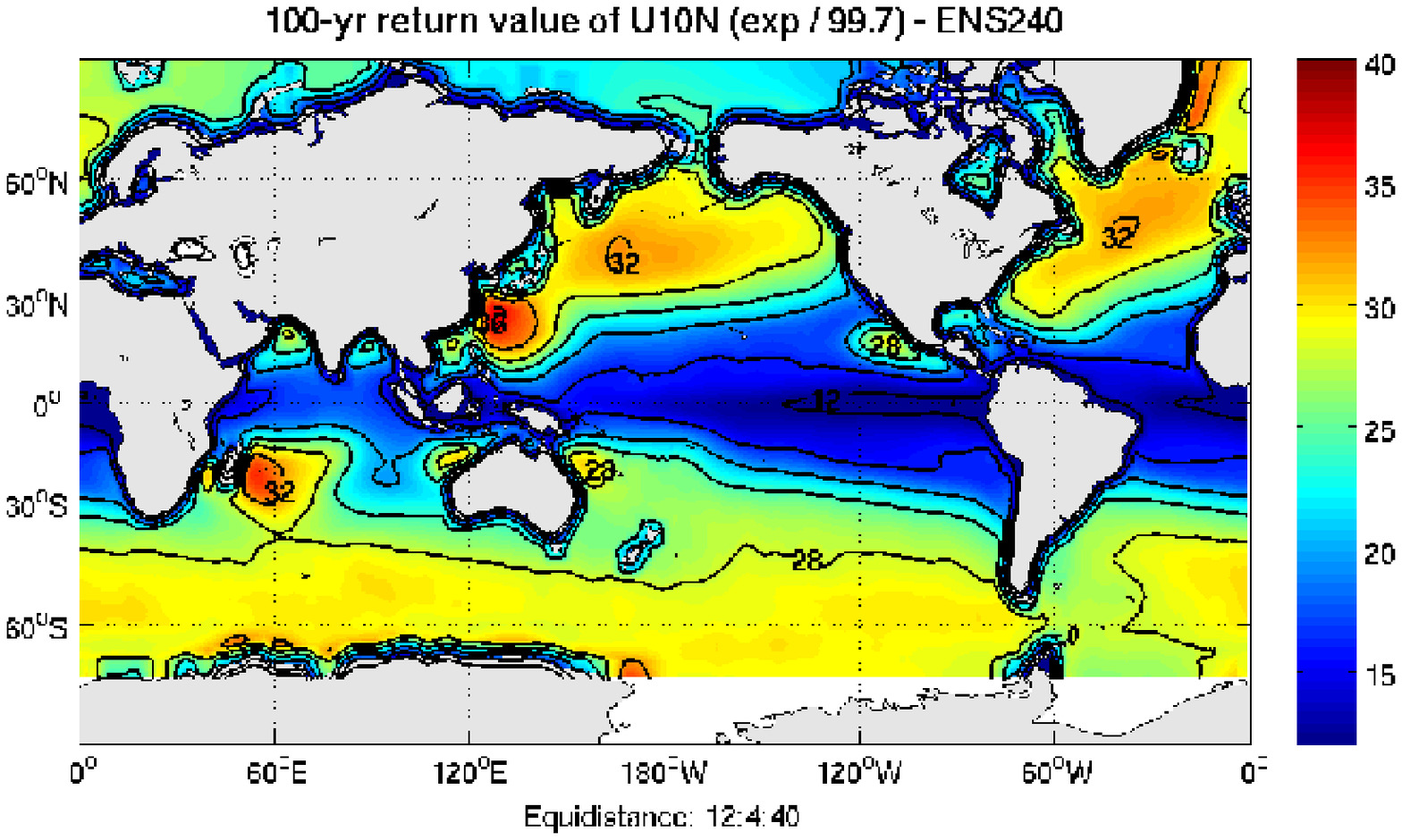}\\
(b)\includegraphics[scale=0.5]{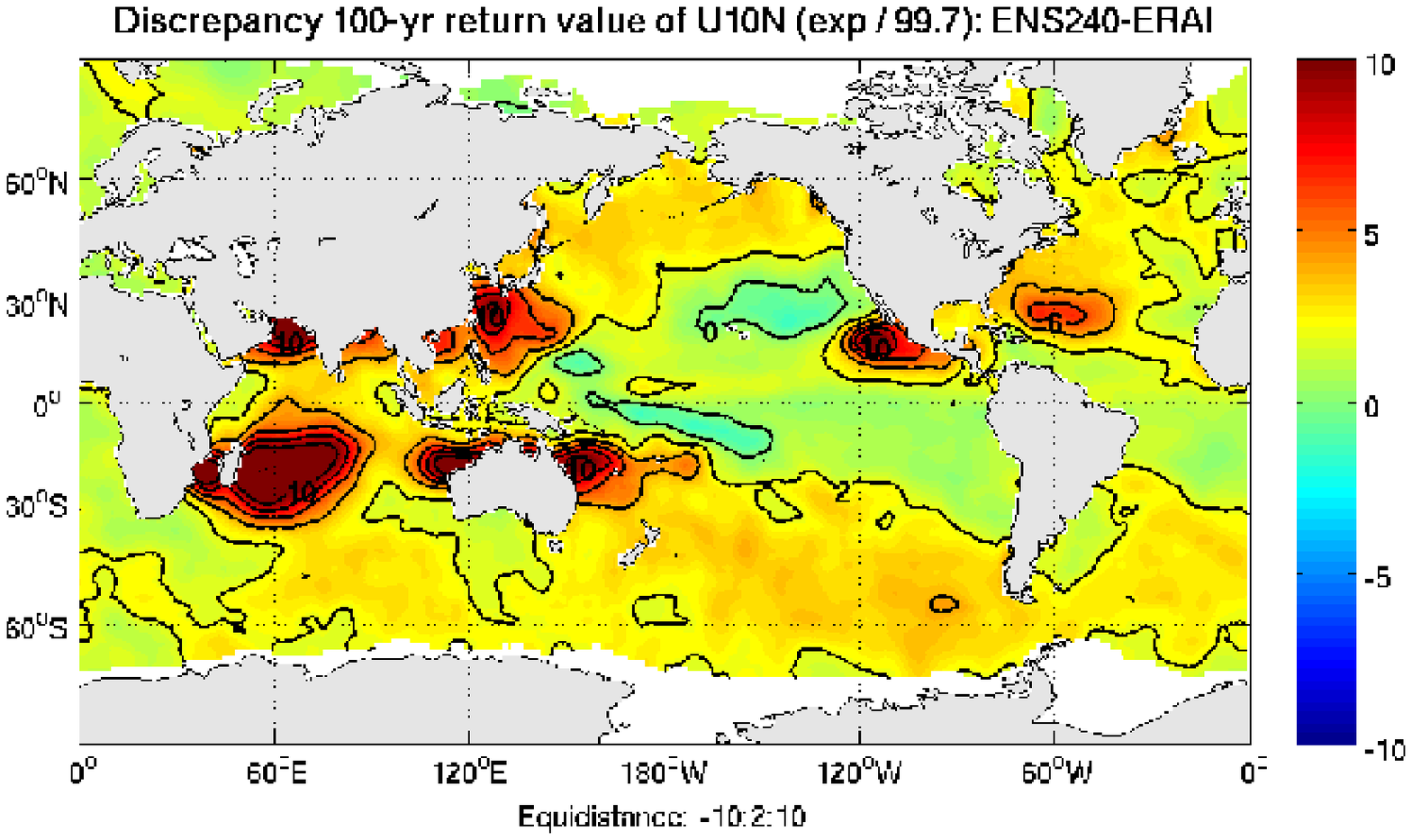}\\
(c)\includegraphics[scale=0.5]{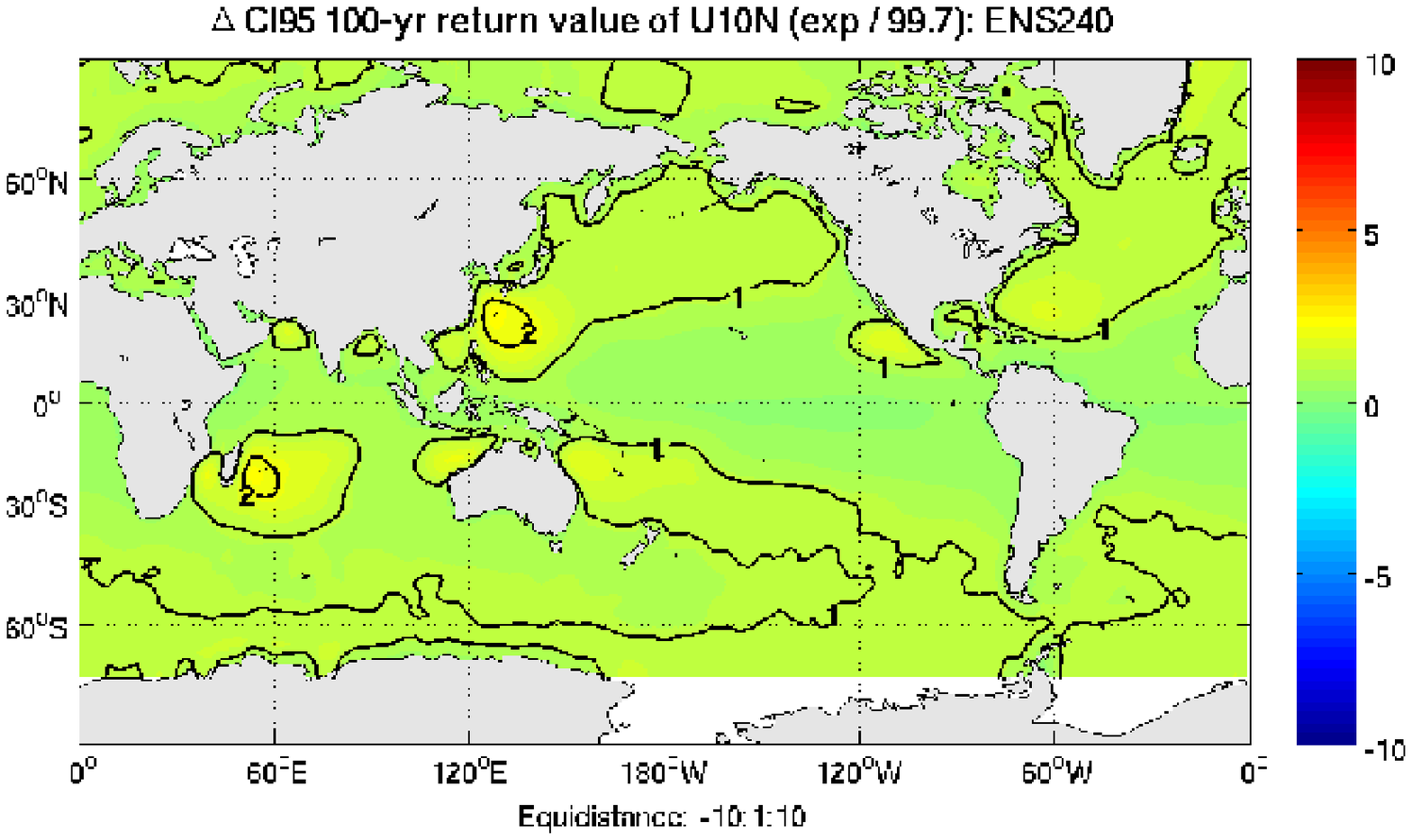}\\
(d)\includegraphics[scale=0.5]{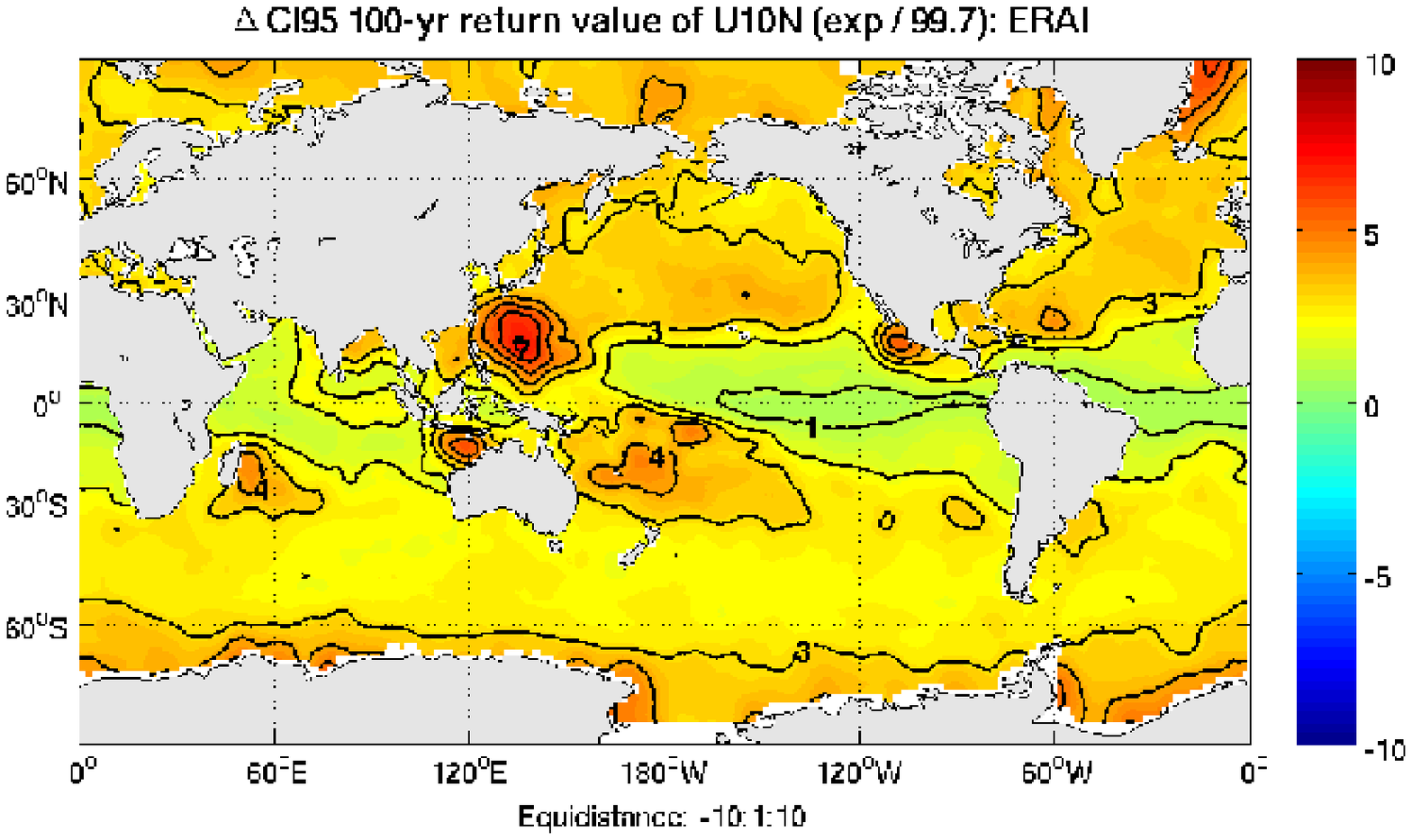}
\caption{10-m wind speed 100-yr return values, $U_{100}$ [m s$^{-1}$]. 
Panel a: ENS estimate. Panel
b: Difference between ENS and ERA-I. Panels c-d: width of 95\% confidence
intervals for ENS240 and ERA-I, respectively.}
\label{fig:U100} 
\end{center} 
\end{figure}

\begin{figure}[h]
\begin{center}
(a)\includegraphics[scale=0.5]{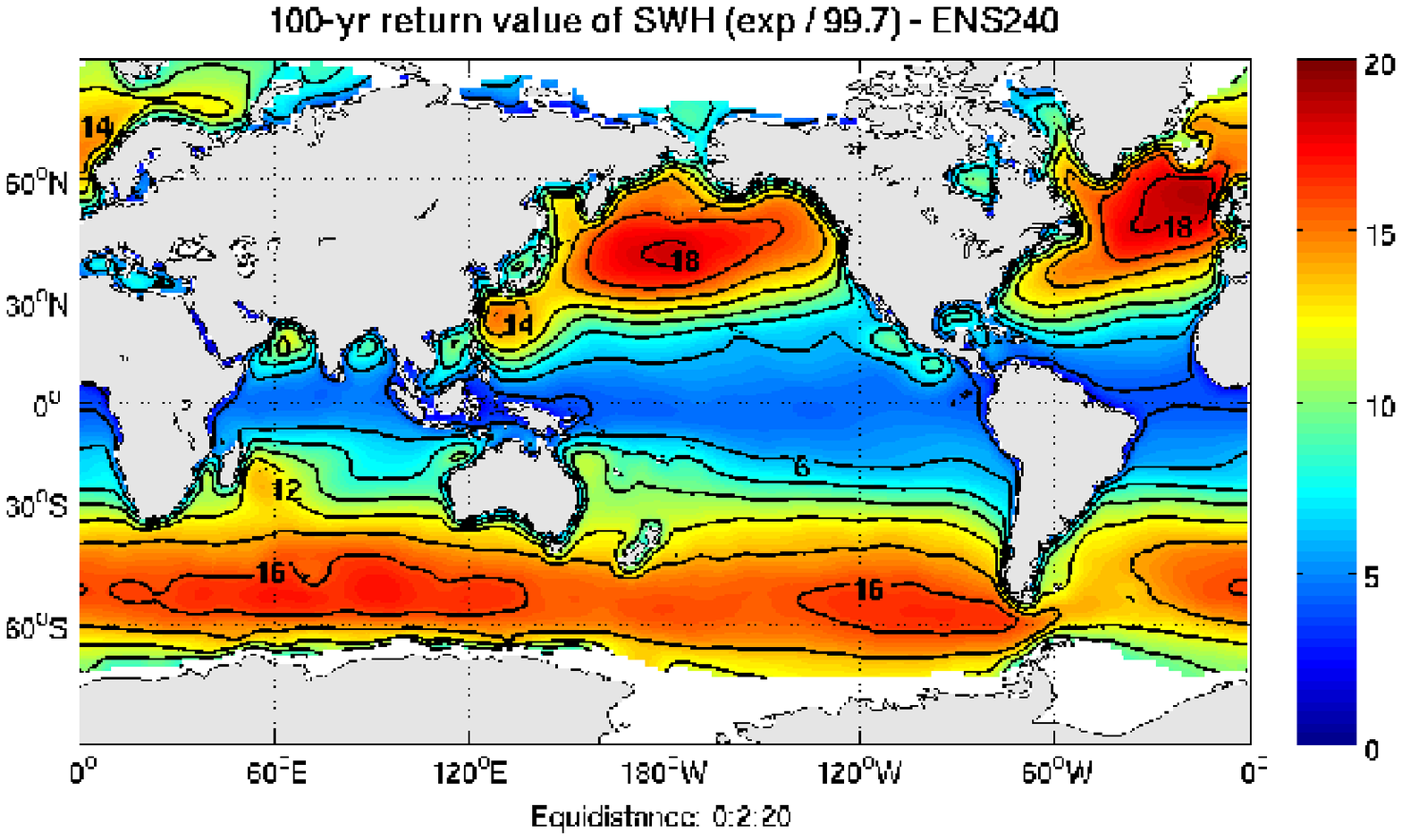}\\
(b)\includegraphics[scale=0.5]{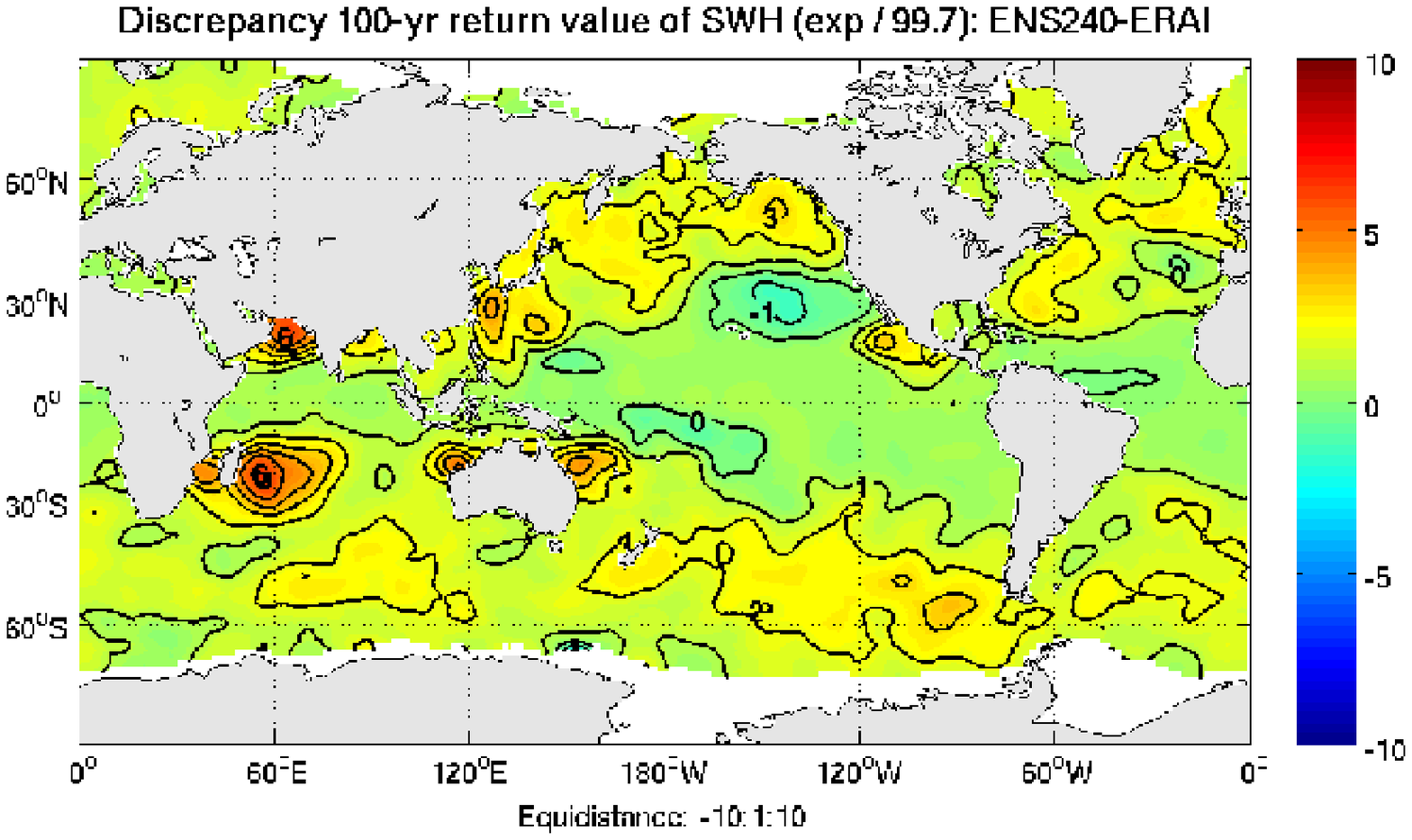}\\
(c)\includegraphics[scale=0.5]{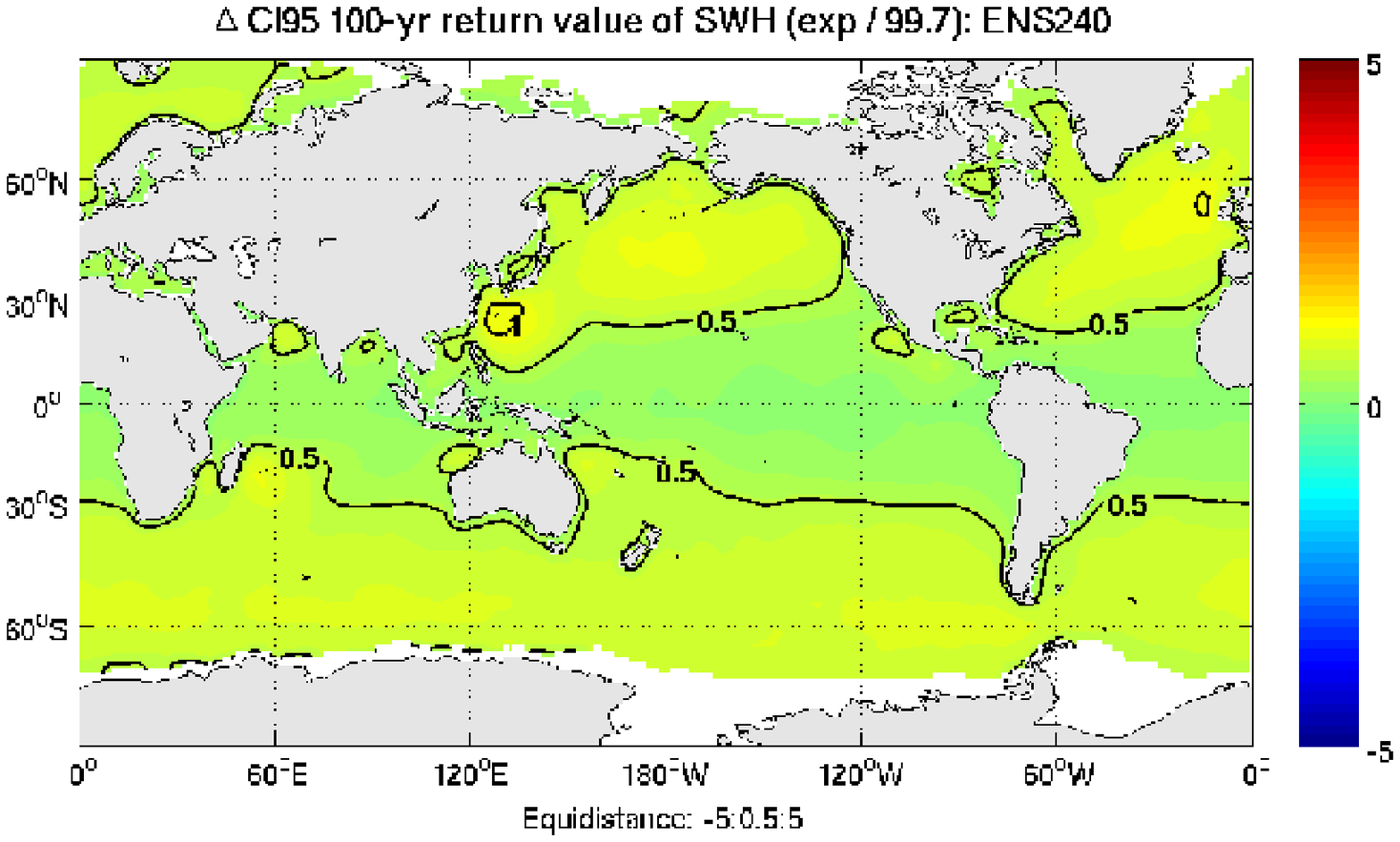}\\
(d)\includegraphics[scale=0.5]{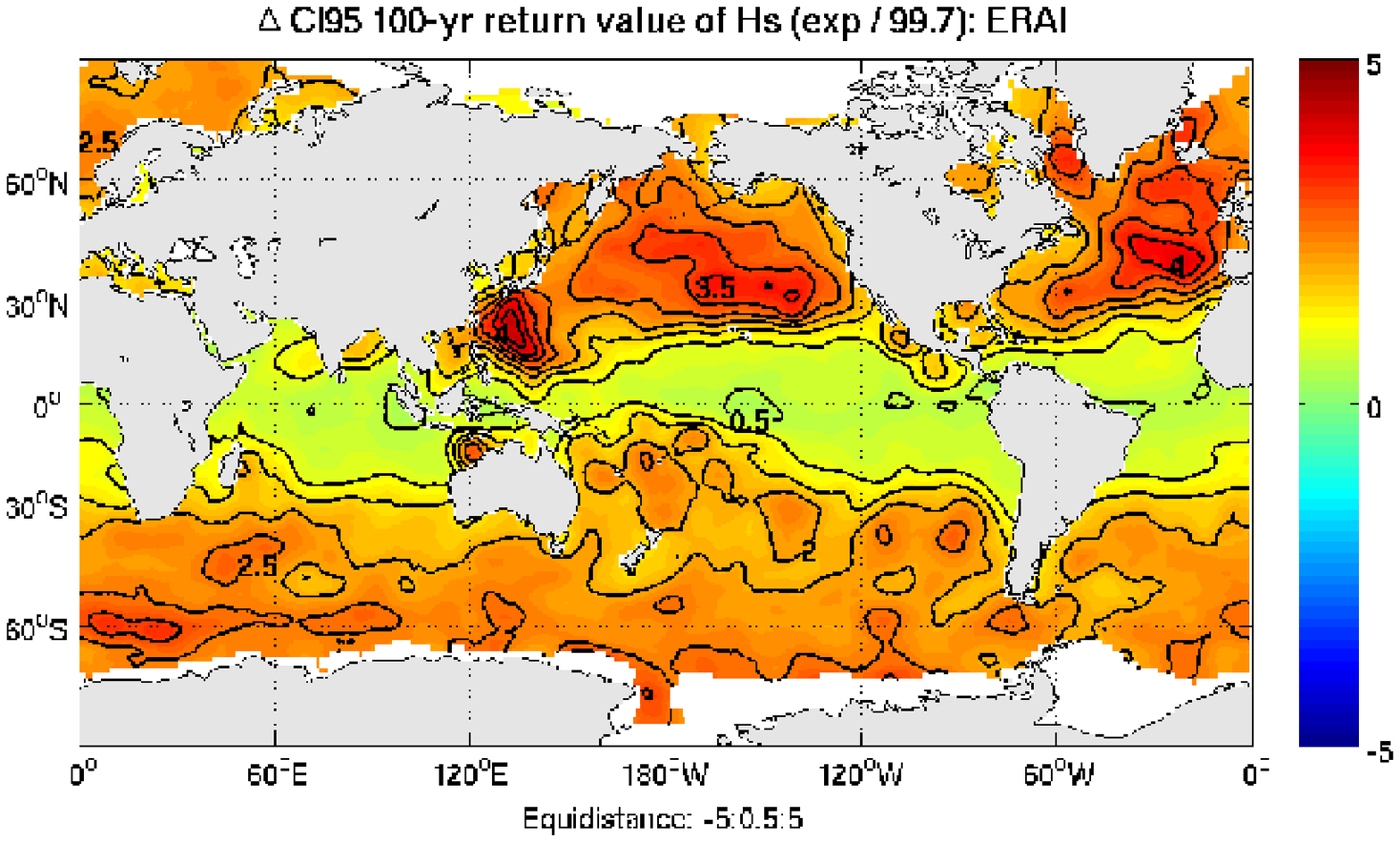}
\caption{
100-yr return values of significant wave height, $H^\mathrm{ENS}_{100}$
[m]. Panel a: ENS estimate. Panel b: Difference between ENS and ERA-I.
Panels c-d: Width of the 95\% confidence intervals for ENS240 and ERA-I,
respectively.}
\label{fig:h100} 
\end{center} 
\end{figure}

The 100-yr return estimate of 10-m wind speed, \Ur, with 95\% confidence
intervals, are shown in \Fig{U100}. The most salient feature is the striking
difference between ERA-I and ENS240 in the subtropics ($>10$ m~s$^{-1}$
difference, see panels b in \Fig{U100} and \Fig{h100}. This is clearly
related to tropical cyclone activity (see e.g. \citealt{oouchi06} for an
overview of geographical distribution of tropical cyclones) Although ENS is
still far from capturing the strength of tropical cyclones, it represents a
substantial improvement over ERA-I.  ENS240 also yields significantly higher
return values ($> 2$ m~s$^{-1}$ difference) in the extratropics (Panel b of
Fig~\ref{fig:U100}).

The same general features are found for the 100-yr return estimate of
the significant wave height, \Hr (\Fig{h100}). In the extratropics we find
differences in excess of 2 m, while in regions of the tropics and subtropics
with high tropical cyclone activity the differences exceed 6 m (east of
Madagascar and in the Arabian Sea in particular).

The ensemble forecasts represent the equivalent of about 229 years, and under
this assumption the confidence intervals are reduced dramatically compared with
the 30 years of ERA-I data. Clearly, as pointed out by \citet{bre13b}, the model
bias persists, but the uncertainty under the assumptions are much lower for the
ensemble data set than for ERA-I. This is clearly seen by comparing Panels c-d
in \Fig{U100} and \Fig{h100}.

Thus it may seem that throughout the extratropics ERA-I underestimates the
100-yr return values for wind speed and wave height by about 10\%, while in
the regions with tropical cyclones the underestimation reaches 25\%. Since
the return values are computed from coarse resolution model simulations,
the ENS240 estimates for the subtropics will be biased low and should be
considered low-end brackets of the real return estimates.

Our \Hr estimates are broadly geographically consistent with previous
estimates of the return values of wind speed and wave height by
\citet{caires05b,sterl05}, based on the ERA-40 \citep{upp05} reanalysis.
However, \citet{caires05b} report as much as 7 m higher wave heights in the
storm tracks in the North Atlantic and the North Pacific.  \citet{vinoth11}
aggregated 30 years of satellite altimeter wind and wave observations. Of
the various extreme value distributions fitted to their data they concluded
that the initial distribution method (IDM) gave the most reliable fit to
the upper percentiles of buoy observations. Their estimates are generally
much higher in the extratropics, and typically 2-4 m higher than ENS240
in the North Atlantic and the North Pacific.  No confidence intervals were
provided by \citet{caires05b} although estimates for different decades of
ERA-40 were computed.  \citet{vinoth11} likewise offer no estimates of the
confidence intervals, but judging by the large spread between the different
methods employed it is likely to be high.

\begin{figure}[h]
\begin{center}
(a)\includegraphics[scale=0.5]{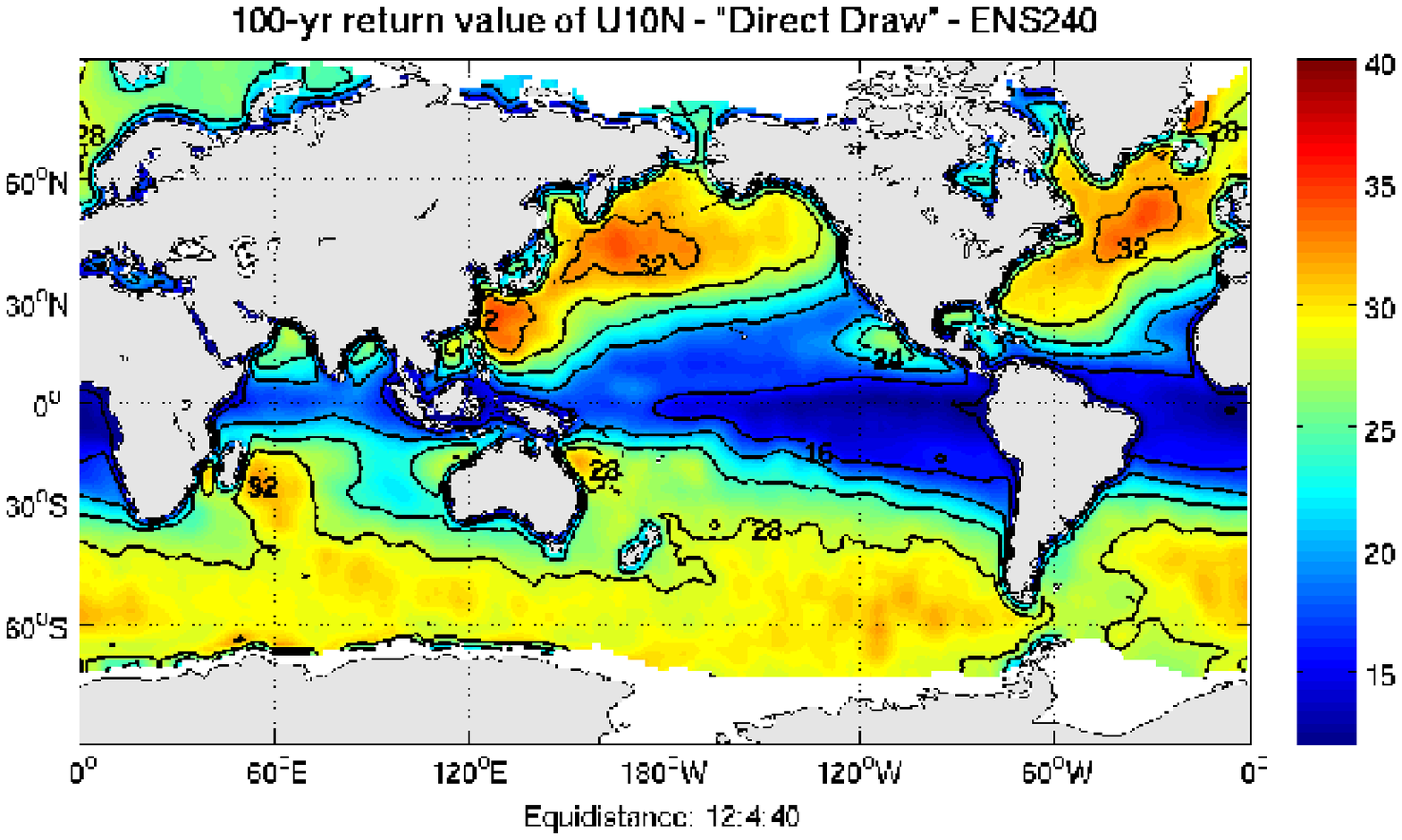}\\
(b)\includegraphics[scale=0.5]{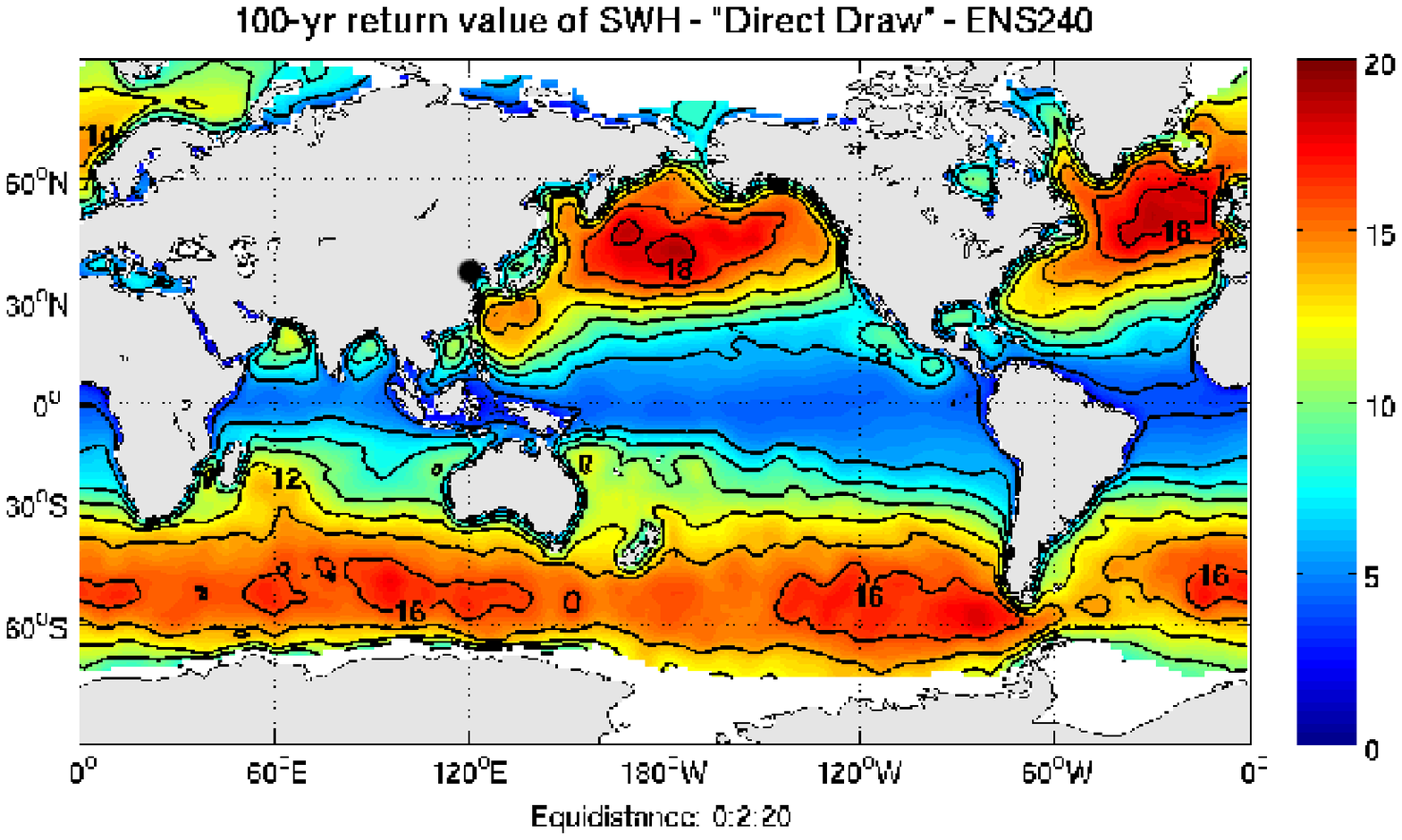}
\caption{Direct return value estimates (DRE). 
Panel a: Direct return estimates of the 100-yr wind speed, $U^\mathrm{DRE}_{100}$ [m s$^{-1}$]. 
Panel b: Similar for significant wave height, $H^\mathrm{DRE}_{100}$ [m].
The return value estimates are 
very similar to those estimated using the exponential distribution.}
\label{fig:dre}
\end{center} 
\end{figure}

To investigate the impact that the exponential fit has on the return estimates
we have computed the non-parametric direct return estimates outlined in
\Sec{methods}. As seen in \Fig{dre} the results are very similar to the
return values computed using the exponential distribution in \Fig{U100}
and \Fig{h100}.

\section{Conclusion}
Return value estimation based on very large aggregates of ensemble forecasts at
advanced lead times was first reported by \citet{bre13b} for wave height in the
Norwegian Sea and the North Sea. We have extended the analysis here to produce
global maps of return values for wind speed and wave height. We find that the
ensemble yields estimates of wave height and wind speed that are signifcantly
higher than ERA-I, but much lower than the estimates reported by
\citet{caires05b} and \citet{vinoth11}. The upper percentiles show good
agreement with buoys and the ENVISAT altimeter. The confidence intervals for
ENS240 are much narrower than for ERA-I due to the much larger data sets (see
Figs \ref{fig:U100} and \ref{fig:h100}). We note that there is substantially
more tropical cyclone activity in the ensemble at long lead times than in ERA-I, which
seems to correspond better with ENVISAT altimeter observations of wind speed and
wave height. However, it is clear that the model is still too coarse to
realistically model wind speed maxima around tropical cyclones and the estimates
for the subtropics are likely to be biased low (but less so than ERA-I). The
return values found in the extratropics seem reasonable and represent a valuable
addition to previous estimates, especially given the much narrower confidence
intervals.

\section*{Acknowledgment}
  This work has been supported by the Research Council of Norway through
  the project ``Wave Ensemble Prediction for Offshore Operations'' (WEPO,
  grant no 200641) and through the European Union FP7 project MyWave (grant
  no 284455). This study has also been part of a PhD program partially funded
  by the Norwegian Centre for Offshore Wind Energy (NORCOWE) for OJA.

{\clearpage}
\bibliography{/home/rd/diob/Doc/TeX/Bibtex/BreivikAbb,/home/rd/diob/Doc/TeX/Bibtex/Breivik} \newpage

\end{document}